\documentclass[universe,review,accept,oneauthor, pdftex,10pt,a4paper]{mdpi}

\lastpage{23}

\doinum{10.3390/universe2020008}
\pubvolume{2}
\pubyear{2016}

\externaleditor{Academic Editors: Lorenzo Iorio and Elias C. Vagenas}
\history{Received: 4 January 2016; Accepted: 11 May 2016; Published: }

\usepackage{amssymb}

\usepackage{amsmath}
\usepackage{amsfonts}
\usepackage{soul}
\usepackage{epstopdf}
\usepackage{attrib}

\newcommand{\be}{\begin{equation}}
\newcommand{\ee}{\end{equation}}
\def\bq{\begin{eqnarray}}
\def\eq{\end{eqnarray}}

\def\simgt{\,\hbox{\lower0.6ex\hbox{$\sim$}\llap{\raise0.6ex\hbox{$>$}}}\,} 
\def\simlt{\,\hbox{\lower0.6ex\hbox{$\sim$}\llap{\raise0.6ex\hbox{$<$}}}\,} 

\Title{Einstein and Beyond: A Critical Perspective on General Relativity}

\Author{Ram Gopal Vishwakarma}

\address[1]{Unidad Acad\'{e}mica de Matem\'{a}ticas,  Universidad Aut\'{o}noma de Zacatecas, 
  {Zacatecas}, ZAC, C.P. 98000, Mexico; E-Mail: vishwa@uaz.edu.mx}

\abstract{An alternative approach to Einstein's theory of General Relativity (GR) is reviewed, which is motivated by a range of serious
theoretical issues inflicting the theory, such as the cosmological constant problem, presence of non-Machian solutions, problems related with the energy-stress tensor $T^{ik}$ and unphysical solutions.\\
 The new approach emanates from a critical analysis of these problems, providing a novel insight that
the matter fields, together with the ensuing gravitational field, are already present inherently in the spacetime without taking recourse to $T^{ik}$.
 Supported by numerous evidences, the new insight  revolutionizes our views on the representation of the source of gravitation and establishes the spacetime itself as the source, which becomes crucial for understanding the unresolved issues in a unified manner. This leads to a new paradigm in GR by establishing equation $R^{ik}=0$ as the field equation of gravitation plus inertia in the very presence of matter.}


\begin{document}


\section{Introduction}

The year 2015 marks the centenary of the advent of Albert Einstein's theory of General Relativity (GR), which constitutes  the current description of gravitation in modern physics.
It is undoubtedly one of the towering theoretical achievements of 20th-century physics, which is recognized as an intellectual achievement par excellence. 

Einstein first revolutionized, in 1905, the concepts of absolute space and absolute time by superseding them with a single four-dimensional spacetime fabric, only which had an absolute meaning. He discovered this in his theory of Special Relativity (SR), which he formulated by postulating that the laws of physics are the same in all non-accelerating reference frames and the speed of light in vacuum never changes. He then made a great leap from SR to GR through his penetrating insight that the gravitational field in a small neighborhood of spacetime is indistinguishable from an appropriate acceleration of the reference frame (principle of equivalence), and hence gravitation can be added to SR (which is valid only in the absence of gravitation) by generalizing it for the accelerating observers. This leads to a curved spacetime.

This dramatically revolutionized the Newtonian notion of gravitation as a force by heralding that gravitation is a manifestation of the dynamically curved spacetime created by the presence of matter. The principle of general covariance (the laws of physics should be the same in all coordinate systems, including the accelerating ones) then suggests that the theory must be formulated by using the language of tensors. 
This leads to the famous Einstein's equation\footnote{Here, as usual,  $g^{ik}$  is the contravariant form of the metric tensor $g_{ik}$ representing the spacetime geometry, which is defined by $ds^2=g_{ik}dx^i dx^k$. $R^{ik}$ is the Ricci tensor defined by $R^{ik}=g_{hj} R^{hijk}$ in terms of the Riemann tensor $R^{hijk}$. $R=g_{ik}R^{ik}$ is the Ricci scalar and $G^{ik}$  the Einstein tensor.  $T^{ik}$ is the energy-stress tensor of matter (which can very well absorb the cosmological constant or any other candidate of dark energy). $G$ is the Newtonian constant of gravitation and $c$ the speed of light in vacuum. The Latin indices range and sum over the values 0,1,2,3 unless stated otherwise.}
\be
G^{ik} \equiv R^{ik} - \frac{1}{2} g^{ik} R = - \frac{8\pi G}{c^4}T^{ik}, \label{eq:EinsteinEq}
\ee
which represents how geometry, encoded in the left hand side (which is a function of the spacetime curvature),  behaves in response to matter encoded in the energy-momentum-stress tensor $T^{ik}$. 
This, in a sense, completes the identification of gravitation with geometry. 
It turns out that the spacetime geometry is no longer a fixed inert background, rather it is a key player in physics, which  acts on matter and can be acted upon. This constitutes a profound paradigm shift.

The theory has made remarkable progress on both, theoretical and observational fronts \cite{progress, Will, Turyshev, Ashtekar, paddy}. It is remarkable that, born a century ago out of almost pure thought, the theory has managed to survive extensive experimental/observational scrutiny and describes accurately all gravitational phenomena ranging from the solar system to the largest scale - the Universe itself.
Nevertheless, a number of questions remain open. On the one hand, the theory requires the dark matter and dark energy - two largest contributions to $T^{ik}$ - which have entirely mysterious physical origins and do not have any non-gravitational or laboratory evidence.
On the other hand, the theory suffers from profound theoretical difficulties, some of which are reviewed in the following.
 Nonetheless, if a theory requires more than  95\% of ``dark entities" in order to describe the observations, it is an alarming signal for us to turn back to the very foundations of the theory, rather than just keep adding epicycles to it.  

Although Einstein, and then others, were mesmerized by the `inner consistency' and elegance of the theory,  many theoretical issues were discovered even during the lifetime of Einstein which were not consistent with the founding principles of GR.
In the following, we provide a critical review of the historical development of GR and some ensued problems, most of which are generally ignored or not given proper attention as they deserve.
This review will differ from the conventional reviews in the sense that unlike most of the traditional reviews, it will not recount a well-documented story of the discovery of GR, rather it will focus on some key problems which insinuate an underlying  new insight on a geometric theory of gravitation, thereby providing a possible way out in the framework of GR itself.

\section{Issues Warranting Attention: Mysteries of the Present with Roots in the Past}

\medskip
\noindent
$\blacktriangleright$ {\bf Mach's Principle:} 
Mach's principle\footnote{The name ``Mach's principle'' was coined by Einstein for the general inspiration that he found in Mach's works on mechanics \cite{Mach}, even though the principle itself was never formulated succinctly by Mach himself.}, akin to the equivalence principle, was the primary motivation and guiding principle for Einstein in the formulation of GR. 
Though in the absence of a clear statement from Ernst Mach, there exist a number of formulations of Mach's principle, in essence the principle advocates to shun all vestiges of the unobservable absolute space and time of Newton in favor of the directly observable background matter in the Universe, which determines its geometry and the inertia of an object.  

As the principle of general covariance (non-existence of a privileged reference frame) emerges just as a consequence of Mach's  denial of absolute space,
Einstein expected that his theory would automatically obey Mach's principle. However, it turned out not to be so, as there appear several anti-Machian features in GR.  According to Mach's principle, the presence of a material background is essential for defining motion and
 a meaningful spacetime geometry.  This means that an isolated object in an otherwise empty
Universe should not possess any inertial properties. But this is clearly violated by the Minkowski solution, which possesses timelike geodesics and a well-defined notion of inertia in the total absence of $T^{ik}$. 
Similarly, the cosmological constant also violates Mach's principle (if it does not represent the vacuum energy, but just a constant of nature - as is believed by some authors) in the sense in which the geometry should be determined completely by the mass distribution.
In the same vein, 
there exists a class of singularity-free curved solutions, which admit Einstein's equations  in the absence of $T^{ik}$.
Further, a global rotation, which is not allowed by Mach's principle (in the absence of an absolute frame of reference), is revealed in the G$\ddot{o}$del solution \cite{Godel}, which describes a universe with a uniform rotation in the whole spacetime.

After failing to formulate GR in a fully Machian sense, Einstein himself moved away from Mach's principle in his later years. Nevertheless, the  principle continued to attract a lot of sympathy due to its aesthetic appeal and enormous impact, and it is widely believed that a viable theory of gravitation must be Machian.
Moreover, the consistency of GR with SR, which too abolishes the absolute space akin to Mach's principle, also persuades us that GR must be Machian.
This characterization has however remained just a wishful thinking.

\medskip
\noindent
$\blacktriangleright$ {\bf Equivalence Principle:} The equivalence principle - the physical foundation of any metric theory of  gravitation - first expressed by Galileo and later reformulated by Newton, was assumed by Einstein as one of the defining principles of GR.
According to the principle, one can choose a locally inertial coordinate system (LICS) (i.e., a freely-falling one) at any spacetime point in an arbitrary gravitational field such that within a sufficiently small region of the point in question, the laws of nature take the same form as in unaccelerated Cartesian coordinate systems in the absence of gravitation \cite{WeinbergBook}.
As has been mentioned earlier, this equivalence of gravitation and accelerated reference frames paved the way for the formulation of GR.
Since, the principle rests on the conviction that the equality of the gravitational and inertial mass is exact \cite{WeinbergBook, Einstein-Grossmann}, one expects the same to hold in GR solutions.
 However, the
 inertial and the (active) gravitational mass have remained unequal in general. For instance, for the case of $T^{ik}$ representing a perfect fluid 
\be
T^{ik}=(\rho +p) u^i u^k-p g^{ik},\label{eq:emtensor}
\ee
various solutions of equation (\ref{eq:EinsteinEq}) indicate that the inertial mass density (= passive gravitational mass density) $=(\rho+p)/c^2$, while the active gravitational mass density $=(\rho +3 p)/c^2$, where $\rho$ is the energy density of the fluid (which includes all the sources of energy of the fluid except the gravitational field energy) and $p$ its pressure.
 The binding energy of the gravitational field is believed to be responsible for this discrepancy. However, why the contributions from the gravitational energy to the different masses are not equal, has remained a mystery.

\medskip
\noindent
$\blacktriangleright$ {\bf $T^{ik}$ \& Gravitational energy:}
 Appearing as the `source term' in equation (\ref{eq:EinsteinEq}), $T^{ik}$ is expected to include all the inertial and gravitational aspects of matter, i.e. all the possible sources of gravitation. However, this requirement does not seem to be met on at least two counts. Firstly, $T^{ik}$ fails to support, in a general spacetime with no symmetries, an unambiguous definition of angular momentum, which is a fundamental and unavoidable characteristic of matter, as is witnessed from the subatomic to the galactic scales. 
While a meaningful notion of the angular momentum in GR always needs the introduction of some additional structure in the form of symmetries, quasi-symmetries, or some other background structure, it can be unambiguously defined only for isolated systems  \cite{M&AM, MTW}.

Secondly, $T^{ik}$ fails to include the energy of the gravitational field, which too  gravitates.
Einstein and Grossmann emphasized that, {\it akin to all other fields, the gravitational field must also have an energy-momentum tensor which should be included in the `source term'} \cite{Einstein-Grossmann}.
However, after failing to find a  tensor representation of the gravitational field, Einstein then commented that {\it ``there may very well be gravitational fields without stress and energy density"} \cite{Einstein1918} and finally
admitted that {\it ``the energy tensor can be regarded only as a provisional means of representing matter''} \cite{Einstein1922}. Alas, a century-long dedicated effort to discover a unanimous formulation of the energy-stress tensor of the
gravitational field, has failed\footnote{
It can be safely said that despite the century-long dedicated efforts of many luminaries, like Einstein, Tolman, Papapetrou, Landau-Lifshitz, M$\ddot{o}$ller and Weinberg, the attempts to discover a unanimous formulation of the gravitational field energy has failed due to the following three reasons: (i) the  non-tensorial character of the energy-stress `complexes' (pseudo tensors) of the gravitational field; (ii) the lack of a unique agreed-upon formula for the gravitational field pseudo tensor in view of various formulations thereof, which may lead to different distributions even in the same spacetime background. Moreover, a pseudo tensor, unlike a true tensor, can be made to vanish at any pre-assigned point by an appropriate transformation of coordinates, rendering its status rather nebulous; (iii)  according to the equivalence principle, the gravitational energy cannot be localized.}, 
concluding that a proper energy-stress tensor of the gravitational field does not exist.
Since then, neither Einstein nor anyone else has been able to discover the true form of $T^{ik}$, although it is at the heart of the current efforts to reconcile GR with quantum mechanics.

It is an undeniable fact that the standards of $T^{ik}$, in terms of elegance, consistency and mathematical completeness, do not match the vibrant geometrical side of equation (\ref{eq:EinsteinEq}), which is determined almost uniquely  by pure mathematical requirements. Einstein himself conceded this fact when he famously remarked:
``{\it GR is similar to a building, one wing of which is made of fine marble, but the other wing of which is built of low grade wood.}" It was his obsession that attempts should be directed to convert the `wood' into `marble'.

The doubt envisioned by Einstein about representing matter by  $T^{ik}$, is further strengthened by a recent study which discovers some surprising inconsistencies and paradoxes in the formulation of the energy-stress tensor of the  matter fields, concluding that the formulation of  $T^{ik}$ does not seem consistent with the geometric description of gravitation \cite{vishwaApSS2}. 
This is reminiscent of the view expressed about four decades ago by J. L. Synge, one of the most distinguished mathematical physicists of the 20th Century: {\it ``the concept of energy-momentum} (tensor) {\it is simply incompatible with general relativity"} \cite{Cooperstock} (which though may seem radical from today's mainstream perspective).

\medskip
\noindent
$\blacktriangleright$ {\bf Unphysical Solutions:}
Since its very inception, GR  started having observational support which substantiated the theory. Its predictions have been well-tested in the limit of the weak gravitational field in the solar system, and in the stronger fields present in the systems of binary pulsars. This has been done through two solutions - the Schwarzschild and Kerr solutions.

However, there exist many other `vacuum' solutions of equation (\ref{eq:EinsteinEq}) which are considered {\it unphysical}, since they represent curvature in the absence of any conventional source.  The solutions falling in this category are the de Sitter solution,  Taub-NUT Solution, Ozsv$\acute{a}$th-Sch$\ddot{u}$cking solution and two newly discovered \cite{pramana, IJGMMP} solutions (given by equations (\ref{eq:int-kerr}) and (\ref{eq:EneDen}) in the following)\footnote{Another solution, which falls in this category, is the G$\ddot{o}$del solution which admits closed timelike-curves and hence permits a possibility to travel in the past, violating the concepts of causality and  creating paradoxes: ``what happens if you go back in the past and kill your father when he was a baby!''}.
Hence the theory has been supplemented by additional `physical grounds' that is used to exclude otherwise exact solutions of Einstein's equation.

This situation is very reminiscent of what
Kinnersley wrote about the GR solutions, {\it ``most of the known exact solutions describe situations which are frankly unphysical"} \cite{Kinnerley}.
This is however misleading because not only does it reject {\it a priori} majority of the exact solutions claiming `unphysical' and `extraneous', but also  mars the general validity of the theory and introduces an element of subjectivity in it. Perhaps we fail to interpret a solution correctly and pronounce it unphysical because the interpretation is done in the framework of the conventional wisdom, which may not be correct \cite{vishwaApSS2, FrontPhys}.

\medskip
\noindent
$\blacktriangleright$ {\bf Interior Solutions:}
As mentioned earlier, GR successfully describes the gravitational field outside the Sun in terms of the Schwarzschild (exterior) and Kerr solutions.  Nevertheless, the theory has not been that successful in describing the interior of a massive body.

Soon after discovering his famous and successful (exterior) solution (with $T^{ik}=0$), Schwarzschild discovered another solution of equation (\ref{eq:EinsteinEq}) (with a non-zero $T^{ik}$) representing the interior of a static, spherically symmetric non-rotating massive body, generally called the Schwarzschild interior solution. After then many other, similar interior solutions have been discovered with different matter distributions.  
It however appears that the picture the conventional interiors provide is not conceptually satisfying.
For example, the Schwarzschild-interior solution assumes  a static sphere of matter consisting of an incompressible perfect fluid of constant density (in order to obtain a mathematically simple solution). Hence the solution turns out to be unphysical, since the speed of sound $=c\sqrt{dp/d\rho}$ becomes infinite  in the fluid with a constant density $\rho$ and a variable pressure $p$. 

The Kerr solution, representing the exterior of a rotating mass, has remained unmatched to any known 
non-vacuum solution that could represent the interior of a rotating mass. It seems that we have been searching for the interior solutions in the wrong place \cite{IJGMMP}.

\medskip
\noindent
$\blacktriangleright$ {\bf Dark Matter \& Dark Energy:}
Soon after formulating GR, Einstein applied his theory to model the Universe. At that time, Einstein believed in a static Universe, perhaps guided by his religious conviction that the Universe must be eternal and unchanging. As  equation (\ref{eq:EinsteinEq}) in its original form does not permit a static Universe, he inserted a term - the famous `cosmological constant $\Lambda$, to force the equation to predict a static Universe. However, it was realized later that this gave an unstable Universe. It was then realized that a naive prediction of equation (\ref{eq:EinsteinEq}) was an expanding Universe, which was subsequently found consistent with the observations. Realizing this, Einstein retracted the introduction of $\Lambda$ terming it his {\it `biggest blunder'}.

The cosmological constant has however reentered the theory in the guise of the dark energy.
As has been mentioned earlier, in order to explain various observations, the theory requires two mysterious, invisible, and as yet unidentified ingredients - dark matter and dark energy - and $\Lambda$ is the principal candidate of dark energy.

One the one hand, the theory predicts that about $27 \%$ of the total content of the Universe is made of non-baryonic dark matter particles, which should certainly be predicted by some extension of the Standard Model of particles physics. However, there is no indication of any new physics beyond the Standard Model which has been successfully verified at the Large Hadron Collider. Curious discrepancies also appear to exist between
the predicted clustering properties of dark matter and observations on small scales.
Obviously the dark matter has eluded our every effort to bring it out of the shadow.

On the other hand, the dark energy is believed to constitute about $68 \%$ of the total content of the Universe. 
 The biggest mystery is not that  the majority of the content of $T^{ik}$ cannot be seen, but that it cannot be comprehended. 
Moreover, the most favored candidate of dark energy - the cosmological constant $\Lambda$ - poses serious conceptual issues, including the
cosmological constant problem - ``why does $\Lambda$ appear to take such an unnatural value?" That is, ``why is the observed value of the energy associated with $\Lambda$
so small (by a factor of $\approx 10^{-120}$!) compared to its value (Planck mass) predicted by the quantum field theory?" and the coincidence problem - ``why is this observed value so close to the present matter density?". 

The cosmological constant problem in fact arises from a structural defect of the field equation (\ref{eq:EinsteinEq}). While in all non-gravitational physics, the dynamical equations describing a system do not change if we shift the `zero point' of energy, this symmetry is not respected by equation (\ref{eq:EinsteinEq}) wherein all sources of energy, stress appear through $T^{ik}$ and hence gravitate (i.e., affect the curvature). As the $\Lambda$-term can very well be assimilated in $T^{ik}$, adding this constant to equation (\ref{eq:EinsteinEq}) changes the solution. 
It may be noted that no dynamical solution of the cosmological constant problem is possible within the existing framework of GR \cite{Weinberg}.

\medskip
\noindent
$\blacktriangleright$ {\bf Horizon Problem:}
Why does the cosmic microwave background (CMB) radiation look the same in all directions despite the photons were emitted from regions of space failing to be causally connected?
The size of the largest coherent region on the last scattering surface, in which the homogenizing signals passed at sound speed, can be measured in terms of the sound horizon. In the standard cosmology, this  however implies that the CMB ought to exhibit large anisotropies ({\it not isotropy}) for angular scales of the order of $1^{\rm o}$ or larger - a result  contrary to what is observed \cite{WeinbergBook}.  Hence, it seems that the isotropy of the CMB cannot be explained in terms of some physical process operating under the principle of causality in the standard paradigm. 

Inflation comes to the rescue. It is generally believed that inflation made the Universe smooth and left the seeds of structures, on the surface of the last scatter, of the order of the Hubble distance at that time. However, inflation has its own problems either unsolved or fundamentally unresolvable. There is no consensus on which (if any) inflation model is correct, given that there are many different inflation models. A physical mechanism that could cause the inflation is not known, though there are many speculations. There are also difficulties on how to turn off the inflation once it starts - the `graceful exit' problem.

\medskip
\noindent
$\blacktriangleright$ {\bf Flatness Problem:}
In the standard cosmology, the total energy density $\rho$ in the early Universe appears to be extremely fine-tunned to its critical value $\rho_{\rm c}=3H^2/(8\pi G$) (which corresponds to a flat spatial geometry of the Universe, here $H$ is the Hubble parameter). Since $\rho$ departs rapidly from $\rho_{\rm c}$ over cosmic time, even a small deviation from this value would have had massive effects on the nature of the present Universe. For instance, the theory requires $\rho$ at the Planck time to be within 1 part in $10^{57}$ of $\rho_{\rm c}$ in order to meet the observed uncertainties in $\rho$ at present! That is, the Universe was almost flat just after the Big Bang. But how? 

 If a theory predicts a fine-tuned value for some parameter, there should be some underlying physical symmetry in the theory. In the present case however, this appears just an unnatural and ad-hoc assumption in order to reproduce observation. 
Inflation comes to the rescue  again. Irregularities in the geometry were evened out by inflation's rapid accelerated expansion causing space to become flatter and hence forcing $\rho$ toward its critical value, no matter what its initial value was.

However, it should also be mentioned that flatness and horizon problems are not problems of GR. Rather, they are problems concerned with the cosmologist's conception of the Universe, very much in the same vein as was Einstein's conception of a static Universe. 

\medskip
\noindent
$\blacktriangleright$ {\bf Scale Invariance:}
It is well-known that GR, unlike the rest of physics, is not scale invariant in the field equation (\ref{eq:EinsteinEq}) \cite{QSSC}. As scale invariance is one of the most fundamental symmetries of physics, any physical theory, including GR, is desired to be scale invariant.

\section{A New Perspective on Gravity}

Thus, with substantial amount of anomalies, paradoxes and unexplained phenomena, one would question whether the pursued approach to GR is correct. 
 Taken at face value, these problems insinuate that our understanding of gravitation in terms of the conventional GR is grossly incomplete (if not incorrect) and we need yet another paradigm shift.
Perhaps we have misunderstood the true nature of a geometric theory of gravitation because of the way the theory has evolved. 

Science advances more from what we do not understand than by what we do understand.
From a careful re-examination of the above-mentioned problems, a new insight with deeper vision of  a geometric theory of gravitation
emerges, which appears as the missing piece of the theory. It may though appear surprising at first sight that these seemingly disconnected problems can lead to any coherent, meaningful solution. Nevertheless, as we shall see in the following, the analysis 
develops drastic revolutionary changes in our conventional views of GR and offers an enlightened view wherein all the above-mentioned difficulties disappear.

\subsection{Revisiting Mach's Principle}

Guided by the principle of covariance, GR has been formulated in the language of tensors. As the principle of covariance results as a consequence from Mach's principle, one naturally expects the theory to be perfectly Machian, as Einstein did.
Then, why do some of the solutions of GR contradict Mach's philosophy? Perhaps we have missed the real message these solutions want to convey. Particularly, the curious presence of the timelike geodesics and a well-defined notion of inertia in the solutions of (\ref{eq:EinsteinEq}) obtained in the absence of $T^{ik}$ must not be just coincidental and there must be some source.

In order to witness this, let us try to impose the philosophy of Mach on the existing framework of GR by quantifying Mach's principle with a precise formulation in which matter and geometry appear to be in one-to-one correspondence. The key insight is the observation that not only inertia, but also space and time emerge from the interaction of matter. As space is an abstraction from the totality of distance-relations between matter, it follows that the existence of matter (fields) is a necessary and sufficient condition for the existence of spacetime.
This idea can be formulated in terms of the following postulate:

\medskip
\noindent
{\it{\bf Postulate:}  Spacetime cannot exist in the absence of fields}. 

\medskip
\noindent
The postulate posits that, the spacetime is not something to which one can ascribe a separate
existence, independently of the matter fields, and the very existence of the spacetime signifies the presence of the matter (fields). This is very much in the spirit of Mach's principle which implies that the existence of a spacetime structure has any meaning only in the presence of matter, which is bound so tightly to the former that one can not exist without the other. 

Inspired by this, Einstein had envisioned that {\it ``space as opposed to `what fills space', has no separate existence"} \cite{Einstein} though he could not implement it in his field equation (\ref{eq:EinsteinEq}), wherein the `space' (represented in the left hand side of the equation)  and `what fills space' (represented by its right hand side) do have separate existence: as has been mentioned earlier, there exist various meaningful spacetime solutions of equation (\ref{eq:EinsteinEq}) in the total absence of $T^{ik}$. The adopted postulate, on the other hand, emphasizes that the spacetime has no independent existence without a material background, which is present universally regardless of the geometry of the spacetime. 

As the matter field is always accompanied by the ensuing gravitational field and since the latter also gravitates, an important consequence of the adopted postulate is that the geometry of the resulting spacetime should be determined by the net contribution from the two fields. Thus the metric field is entirely governed by considered matter fields, as one should expect from a Machian theory.

\subsection{Fields Without $T^{ik}$: An Inescapable Consequence of Mach's Principle}

The theoretical appeal of the above-described hypothesis is that it is naive, self evident and plausible. But more than that it has potential to shape a theory and gives rise to a new vision of GR with novel, dramatic implications. For instance, it makes a powerful prediction that the resulting theory should not have any bearing on the energy-stress tensor $T^{ik}$ in order to represent the source fields\footnote{The source of curvature in a solution of Einstein's field equation (\ref{eq:EinsteinEq}), in the absence of $T^{ik}$, is conventionally attributed to a singularity. This prescription is however rendered nebulous by the presence of various singularity-free curved solutions of (\ref{eq:EinsteinEq}) in the absence of $T^{ik}$.}. Let us recall that equation (\ref{eq:EinsteinEq}) does admit various meaningful spacetime solutions in the absence of the `source term' $T^{ik}$.

According to the postulate, as fields are present universally in all spacetimes irrespective of their geometry, the flat Minkowskian spacetime should not be an exception and it must also be endowed with the matter fields and the ensuing gravitational field.
Now let us recall that the Minkowski spacetime appears as a solution of Einstein's field equation (\ref{eq:EinsteinEq})
only in the absence of $T^{ik}$, in which case the effective field equation yields
\be
R^{ik}=0.\label{eq:RicciEq} 
\ee
However, if the fields can exist in the Minkowski spacetime (as asserted by the founding postulate) in the absence of $T^{ik}$, they can also exist in other spacetimes in the absence of $T^{ik}$. Hence the requirement of uniqueness of the field equation of a viable theory, dictates that $T^{ik}$ must not be the carrier of the source fields in a theory resulting from the adopted postulate and thus, the canonical equation (\ref{eq:RicciEq}) emerges as the field equation of the resulting theory in the very presence of matter. In fact this is what happens if we accept, at their face value, the implications of Mach's principle applied to GR.

This novel feature GR would acquire - that the spacetime solutions of (\ref{eq:RicciEq}), including the Minkowskian one,  are not devoid of fields -  provides an appealing first principle approach  and a linchpin to  understand various unsolved issues in a unified scheme. It becomes remarkably decisive for the theory on Machianity. It was the earlier-mentioned characteristic of the Minkowski and other solutions of (\ref{eq:RicciEq}) to possess timelike geodesics and a well-defined notion of inertia, which pronounced these solutions  non-Machian, as they are conventionally regarded to represent empty spacetimes. The new insight however renders them perfectly Machian and physically meaningful by bestowing a matter-full dignity on them.
Moreover, this novel feature of the Minkowski solution also explains another so-far unexplained issue: It has been noticed that the Noether current associated with an arbitrary vector field in the Minkowski solution is non-zero in general \cite{Paddy}, which remains unexplained in the conventional `empty' Minkowskian spacetime.

Though the proposed scheme of having matter fields in the absence of $T^{ik}$ may sound surprising and orthogonal to the prevailing perspective, it seems to have many advantages over the conventional approach, as we shall see in the following. The issue is whether it can be made realistic. That is, if equation (\ref{eq:RicciEq}) is claimed to constitute the field equation of a viable theory of gravitation in the very presence of matter, its solutions must possess some imprint of this matter. So, do we have any evidence of such imprints in the solutions of equation (\ref{eq:RicciEq})? The answer is, yes.

\subsection{Evidences of the Presence of Fields in the Absence of $T^{ik}$}

As Mach's principle denies unobservable absolute spacetime in favor of the observable quantities
(the background matter) which determine its geometry, the principle would
expect the source of curvature in a solution to be attributable entirely to some directly observable quantity, such as mass-energy, momentum, angular momentum or their densities. So, if GR is correct and it must be Machian, 
these quantities are expected to be supported by some dimension-full parameters appearing in the curved spacetime solutions in such a way that the parameters vanish as the observable quantities vanish, reducing the solutions to the Minkowskian form. 

Interestingly, it has been shown recently \cite{pramana, IJGMMP} that it is always possible to write a curved solution of (\ref{eq:RicciEq}) in a form containing some dimension-full parameters, which appear in the Riemann tensor generatively and can be attributed to the source of curvature. The study further shows that these parameters can support physical observable quantities such as the mass-energy, momentum or angular momentum or their densities.
For instance, the source of curvature in the Schwarzschild solution
\be
ds^2=\left(1+\frac{K}{r}\right)c^2 dt^2-\frac{dr^2}{(1+K/r)}-r^2d\theta^2-r^2\sin^2\theta ~d\phi^2,\label{eq:Sch}
\ee
can be attributed to the mass $m$ (of the isotropic matter situated at $r=0$) through the parameter $K=-2Gm/c^2$. 
Similarly, the dimension-full parameters present in the Kerr solution can be attributed to the mass and the angular momentum of the source mass; those in the Taub-NUT solution to the mass and the momentum of the source; and the parameters in the Kerr-NUT solution to the mass, momentum and angular momentum \cite{pramana, IJGMMP}.

A remarkable evidence of the presence of fields in the absence of $T^{ik}$ is provided by the Kasner solution which exemplifies that even in the standard paradigm, all the well-known curved solutions of equation (\ref{eq:RicciEq}) do not represent space outside a gravitating mass in an empty space\footnote{It is conventionally believed that only those curved solutions of equation (\ref{eq:RicciEq}) are meaningful which represent space outside some source matter, otherwise the solutions represent an empty spacetime.  However, equation (\ref{eq:RicciEq}) cannot decipher just form the symmetry of a solution that it necessarily belongs to a spacetime structure in an empty space outside a mass, since the same symmetry can also be shared by a spacetime structure inside a matter distribution.}. 
Although the Kasner solution in its standard form does not contain any dimension-full parameter which can be attributed to its curvature, however the solution can be transformed to the form
\be
ds^2=c^2 dt^2- (1+nt)^{2p_1}dx^2- (1+nt)^{2p_2}dy^2
- (1+nt)^{2p_3}dz^2,\label{eq:kasner}
\ee
where $n$ is an arbitrary constant parameter (which is dimension-full) and the dimensionless parameters $p_1$, $p_2$, $p_3$ satisfy
$p_1+p_2+p_3=1=p_1^2+p_2^2+p_3^2.$

A dimensional analysis suggests that in order to meet its natural dimension (which is of the dimension of the inverse of time), the parameter $n$ can support only the densities of the observables energy, momentum or angular momentum and {\it not} the energy, momentum or angular momentum themselves [such that (\ref{eq:kasner}) becomes Minkowskian when the observables vanish]. However, the energy density and the angular momentum density vanish here: while the symmetries of (\ref{eq:kasner}) discard any possibility for the angular momentum density, the energy density disappears as it is canceled by the negative gravitational energy \cite{IJGMMP, Hawking-Milodinow}. That is, the parameter $n$ in  (\ref{eq:kasner}) can be expressed  in terms of the momentum density ${\cal P}$ as $n=\gamma \sqrt{G{\cal P}/c}$, where $\gamma$ is a dimensionless constant. 
This indicates that solution  (\ref{eq:kasner}) results from a (uniform) matter distribution (throughout space) and not from a spacetime outside a point mass as in the cases of the Schwarzschild and Kerr solutions. Thus the Kasner solution represents a homogeneous distribution of matter expanding and contracting anisotropically (at different rates in different directions), which can give rise to a net non-zero momentum density represented through the parameter $n$ serving as  the source of curvature, thus demystifying the solution. 

This new insight on the source of curvature is  authenticated by two new solutions of equation (\ref{eq:RicciEq}) discovered in \cite{pramana, IJGMMP} whose discovery is facilitated by the new insight.
The first solution, whose source of curvature cannot be explained with the conventional wisdom (as it is singularity-free), provides a powerful support to the proposed Machian strategy of representing the source in terms of the dimension-full source-carrier parameters. The solution is given by
\be
ds^2 =\left(1-\frac{\ell^2x^2}{8}\right)c^2dt^2-dx^2-dy^2-\left(1+\frac{\ell^2x^2}{8}\right)dz^2
+\ell x (cdt-dz)dy+\frac{\ell^2x^2}{4}cdt~dz,\label{eq:int-kerr}
\ee
which has been derived by defining the parameter $\ell$ in terms of the angular momentum density ${\cal J}$ via  $\ell=G{\cal J}/c^3$ \cite{pramana}. The fact that the parameter $\ell$ can support only the density of angular momentum and {\it not} the angular momentum itself, asserts  that solution (\ref{eq:int-kerr}) results from a rotating {\it matter distribution} (confined to
$-\frac{2\sqrt{2}}{|\ell|}<x<\frac{2\sqrt{2}}{|\ell|}$) and not from a spacetime outside a point mass as are the cases of the Schwarzschild and Kerr solutions.  This is in perfect agreement with the founding postulate that the fields are not different from the spacetime.
 
Solution (\ref{eq:int-kerr}) as a new solution of field equation (\ref{eq:RicciEq}) is important in its own right. Moreover, it illuminates the so far obscure source of curvature in the well-known Ozsv$\acute{a}$th-Sch$\ddot{u}$cking solution, which would otherwise be in stark contrast with the new strategy in the absence of any free parameter. It has been shown in \cite{pramana} that the Ozsv$\acute{a}$th-Sch$\ddot{u}$cking solution results from (\ref{eq:int-kerr}) by assigning a particular value to the parameter $\ell$.

Following the new insight, another new solution of equation (\ref{eq:RicciEq}) has been discovered recently in \cite{IJGMMP}, whose curvature is supported by the energy density. The solution\footnote{The author recently came to know that solution (\ref{eq:EneDen}) has also been reported in \cite{Giardino}.} is given by
\be
ds^2=\frac{(1+4\mu z^2)}{(1+ \mu r^2)^2}c^2 dt^2 - \frac{ dr^2}{(1+ \mu r^2)^4} - r^2 d\phi^2 - \frac{dz^2}{(1+4 \mu z^2)(1+ \mu r^2)^2},\label{eq:EneDen}
\ee
which represents an inhomogeneous axisymmetric distribution of matter, with the parameter $\mu$ given in terms of the energy density ${\cal E}$ as $\mu=G{\cal E}/c^4$. 
As solution  (\ref{eq:EneDen}) is curved but singularity-free for all finite values of the coordinates, it provides, in the absence of any conventional source there, a strong support to the new strategy of source representation.

\subsection{A New Vision of Gravity in the Framework of GR: Spacetime Becomes a Physical Entity}

What does the presence of these dimension-full parameters we witness in the solutions of the field equation (\ref{eq:RicciEq}) signify? As the physical observable quantities sustained by the parameters - i.e. energy, momentum, angular momentum and their densities - have any meaning only in the presence of matter, the presence of such parameters in the solutions of equation (\ref{eq:RicciEq}) must not be just a big coincidence and at face
value, their ubiquitous presence in the solutions of equation (\ref{eq:RicciEq}) insinuates that fields are
universally present in the spacetime in equation (\ref{eq:RicciEq}).

Not only does this provide a strong support to the founding postulate establishing GR as a Machian theory, but also establishes, on firm grounds,  equation (\ref{eq:RicciEq}) as the field equation of a feasible theory of gravitation in the very presence of fields. More than that, thence emerges a radically new vision of  a geometric theory of gravitation through drastic revolutionary changes in our views on the representation of the source of gravitation, which must be through the geometry and {\it not} through $T^{ik}$. By reconceptualizing our previous notions of spacetime, this constitutes a paradigm shift in GR wherein the spacetime itself becomes a physical entity, we may call it the `emergent matter' in a  relativistic/geometric theory of gravitation. From the ubiquitous presence of fields in all geometries, it becomes clear that there is no empty space solution in the new paradigm, as one should expect from a Machian theory. The same was also envisioned by Einstein (though could not be achieved).

One may wonder how the properties of matter can be incorporated into the dynamical equations of the new theory without taking recourse to $T^{ik}$.
This can be achieved by applying the conservation laws and symmetry principles to the new conviction that  all spacetimes harbor fields, inertial and gravitational, whose net contribution determines their geometry. For instance, by assuming  that the sum of the gravitational and inertial energies in a uniform matter distribution should be vanishing \cite{IJGMMP, Hawking-Milodinow}, it has been shown recently that
the homogeneous, isotropic Universe in the new paradigm leads to the Friedmann equation of the standard  `concordance' cosmology \cite{IJGMMP}.   
This should not be a surprise, as  the Friedmann equation, for dust, can also be derived in Newtonian cosmology or in a kinematic theory (like the Milne model) by using the continuity equation and the Navier-Stokes equation of fluid dynamics \cite{narlikar, vishwa_milne}.

\subsection{Equivalence Principle in the New Perspective}

The perfect equivalence between gravitational and inertial masses, first noted by Galileo and Newton, was more or less accidental. For Einstein however this served as a key to a deeper understanding of inertia and gravitation. From his valuable insight that the kinematic acceleration and the acceleration due to gravity are intrinsically identical, he was able to unearth a hitherto unknown mystery of nature - that gravitation is a geometric phenomenon. 

It however seems that the full implications of the equivalence principle have not yet been appreciated. If gravitation is a geometric phenomenon, then through  the (local) equivalence of gravitation and inertia, the inertia of matter should also be considered geometrical in nature, at least when it appears in a geometric theory of gravitation. A purely geometrical interpretation of gravitation would be impossible unless the gravitational as well as the inertial properties of matter are intrinsically geometrical. This would however have revolutionary implications. Considering $T^{ik}$
(which represents the inertial fields) of a purely geometric origin, equation (\ref{eq:EinsteinEq}) would imply
\be
G^{ik} + \frac{8\pi G}{c^4}T^{ik} \equiv \chi^{ik}=0, \label{eq:EinsteinNewEq}
\ee
where $\chi^{ik}$ appears a tensor of purely geometric origin. This would however be nothing else but the Ricci tensor $R^{ik}$ (with a suitable $g_{ik}$), since the only tensor of rank two having a purely geometric origin (emerging from the Riemann tensor), is the Ricci tensor. That is, equation (\ref{eq:EinsteinNewEq}) would reduce to the field equation (\ref{eq:RicciEq})!
In this way, the consequences of the equivalence principle would be in perfect agreement with the adopted Machian postulate - that spacetime has no separate existence
from matter, i.e., the parameters of the spacetime geometry determine entirely the combined effects of gravitation and inertia.

Therefore, taken together with Mach's principle, the consequence of the equivalence principle -  that the gravitational and inertial fields are entirely geometrical by nature - takes GR to its logical extreme in that the spacetime emerges from the interaction of matter. This reconceptualizes the previous notion of spacetime by establishing it as the very source of gravitation. The matter is in fact more intrinsically related to the geometry than is believed in the conventional GR and all the aspects of matter fields (including the
ensuing gravitational field) are already present inherently in the spacetime geometry. This establishes equation (\ref{eq:RicciEq}) as a competent field equation of gravitation plus inertia. This is well-supported by our observation that while the gravitational field is present in the Schwarzschild, Kerr and Taub-NUT solutions (as these represent the spacetimes outside the source mass), the inertial as well as the gravitational fields are present in solutions (\ref{eq:kasner}),  (\ref{eq:int-kerr}) and (\ref{eq:EneDen}) including the Minkowskian one, which represent matter distribution.

A precise specification of the fields, which are being claimed to be present in the spacetime, is possible only when a precise formulation thereof is available. Nevertheless, in view of the newly gained insight, at least this much can be declared that the matter fields present in the geometry of equation (\ref{eq:RicciEq}) are those which are attempted to be introduced  in equations (\ref{eq:EinsteinEq}) or (\ref{eq:EinsteinNewEq}) via $T^{ik}$ (which has now been absorbed in (\ref{eq:RicciEq})).

\section{A Closer Look at the Conventional 4-Dimensional Formulation of Matter}

Modeling matter by $T^{ik}$ in equation (\ref{eq:EinsteinEq}) has modified at the deepest level the way we used to think about the source of gravitation.
As mass density is the source of gravitation in Newtonian theory, the
energy density was expected to take over this role in the relativistic generalization of
Poisson's equation. To our surprise however, all the ten (independent)
components of $T^{ik}$ become contributing source of gravitation. We need not doubt this novelty, as the new theory,  originated from the innovative ideas, is expected to have innovative features.
But, the way the non-conventional sources appear in the dynamical equations, appears to create inconsistencies and paradoxes, which
 warrants a second look at the relativistic formulation of matter given by  $T^{ik}$.

Everyone will agree that like the conservation of momentum, the conservation of energy of an isolated system is an absolute symmetry of nature and this fundamental principle is expected to be respected by any physical theory.
Nonetheless, the principle is violated in GR in many different situations including the cosmological scenarios (see, for example, \cite{harrison}). The blame rests with the energy of the gravitational field, which has been of an obscure nature and a controversial history, as has been mentioned earlier.
We shall however see that the gravitational energy is not to be blamed for the trouble. This is ascertained beyond doubts in the following analysis by filtering out the gravitational energy from the equations.

\subsection{Problems with $T^{ik}$}

As is well-known, the formulation of the energy-stress tensor $T^{ik}$ given by (\ref{eq:emtensor}) is obtained by first deriving it in the absence of gravity in SR, by considering a  fluid element in a small neighbourhood of an LICS, which exists admittedly at all points of spacetime (by courtesy of the principle of equivalence). Then the expression for the tensor in the presence of gravity is imported, from SR to GR, through a coordinate transformation. 
It would be insightful to reconsider the same LICS to understand the mysterious implications of $T^{ik}$, if any, since the subtleties of gravitation and the gravitational energy disappear locally in this coordinate system. Let us then study the divergence of $T^{ik}$ in the considered  LICS, which is known for describing the mechanical behaviour of the fluid.
Through the vanishing divergence of $G^{ik}$, (\ref{eq:EinsteinEq}) implies that $T^{ij}_{~ ~;j}=0$, which, in the chosen coordinates, reduces to
\be
\frac{\partial T^{ij}}{\partial x^j}=0.\label{eq:consf}
\ee

\bigskip
\noindent
$\bullet$
For the case of a perfect fluid given by (\ref{eq:emtensor}), it is easy to show that equation (\ref{eq:consf}), in the chosen LICS, yields \cite{tolman}
\be
\frac{\partial p}{\partial x}+\frac{\left(\rho+p\right)}{c^2}\frac{du_x}{dt}=0,
\label{eq:mu1}
\ee
for the case $i=1$, where $du_x/dt$ is the acceleration of the considered fluid element in the x-direction. 
As any role of gravity and gravitational energy is absent in this equation, it can be interpreted as the relativistic analogue of the Newtonian law of motion: the fluid element of unit volume, which moves under the action of the force applied by the pressure gradient $\partial p/\partial x$, has got the inertial mass $(\rho+p)/c^2$.
Let us however recall that the term $\rho$ in equation (\ref{eq:emtensor}) includes in it, by definition, not only the rest mass of the individual particles of the fluid but also their kinetic energy, internal energy (for example, the energy of compression, energy of nuclear binding, etc.)
and {\it all other sources of mass-energy} \cite{MTW}.
Therefore, the additional contribution to the inertial mass entering through the term $p$, appears to violate the celebrated law of the conservation of energy.
Though (\ref{eq:mu1}) is usually interpreted as a momentum conservation
equation, however an alternative (but viable) interpretation is not expected to defy the energy conservation.

\bigskip
\noindent
$\bullet$
Similar problems seem to afflict the temporal component of equation (\ref{eq:consf}) for $i=0$, which can be written, following \cite{tolman} as
\be
\frac{d}{dt}(\rho \delta v)+p\frac{d}{dt}(\delta v)=0,\label{eq:mu0}
\ee
where $\delta v$ is the proper volume of the fluid element. The usual interpretation to this equation says: the rate of change in the energy of the fluid element is given in terms of the work done against the external pressure. This seems reasonable at the first sight, but cracks seem to appear in it after a little reflection. The concern, as also noticed by Tolman \cite{tolman}, is that the fluid of a finite size can be divided into similar fluid elements and the same equation (\ref{eq:mu0}) can be applied to each of these elements, meaning that the proper energy $(\rho \delta v)$ of every element is decreasing when the fluid is expanding or increasing when the fluid is contracting. This leads to a paradoxical situation that the sum of the proper energies of the fluid elements which make up an isolated system, is not constant. Tolman overlooked this problem by assuming a possible role of the gravitational energy in it. We however note that no such possibility exists as equation (\ref{eq:mu0}) has been derived in an LICS.

\bigskip
\noindent
$\bullet$
The total energy $E$, including the gravitational energy, of an isolated time-independent fluid sphere comprising of perfect fluid given by (\ref{eq:emtensor}) and occupying  volume $V$ of the 3-space $x^0=$ constant, is given by the Tolman formula \cite{tolman}
\be
E=\int_V (\rho +3p)\sqrt{|g_{00}|}~dV,\label{eq:rmass} 
\ee
which measures the strength of the gravitational field produced by the fluid sphere. The formula is believed to be consistent, for the case of the disordered radiation ($p=\rho/3$), with the observed deflection of starlight (twice as much as predicted by a heuristic argument made in Newtonian gravity), when it passes past the Sun.
Ironically, this expectation is contradicted by the weak-field approximation of the same equation (\ref{eq:rmass}). In a weak field, like that of the Sun, where Newtonian gravitation can be regarded as a satisfactory approximation, equation (\ref{eq:rmass}) can be written, following Tolman (see page 250 of \cite{tolman}), as
$E=\int \rho dV +(1/2c^2)\int \rho \psi dV,$
where $\psi$ is the Newtonian gravitational potential. As $\psi$ is negative, we note that the general relativistic active gravitational mass $E/c^2$ of the gravitating body, here Sun, is obviously less than its Newtonian value  $(1/c^2)\int \rho dV$ and is expected to give a lower value for the gravitational deflection of light than the corresponding Newtonian value! Let us recall that the correct interpretation of the observations of the bending of starlight, when it passes past the Sun, comes from the correct geometry around Sun resulting from the Schwarzschild solution.

\bigskip
\noindent
$\bullet$
As has been mentioned earlier, the (active) gravitational and inertial mass are in general unequal in GR solutions (the discrepancy thereof is supposed to be accounted by the gravitational energy). Thus in an LICS, which nullifies gravitation and hence gravitational mass locally, we expect  a unique value for the mass (density) in the equations. To check this, let us calculate the Tolman integral (\ref{eq:rmass}) in the considered LICS wherein it reduces to
\be
E=\int (\rho +3p)~dV,\label{eq:rmassin} 
\ee
which may now be valid for a sufficiently small volume of the fluid. Surprisingly, we still encounter different unequal values of mass (density) in equations (\ref{eq:mu1}), (\ref{eq:mu0}) and (\ref{eq:rmassin}).  [Equation (\ref{eq:mu0}) can be written alternatively as
$\delta v ~d\rho/dt+(\rho+p)d(\delta v)/dt=0.$]. While equations (\ref{eq:mu1}) and (\ref{eq:mu0}) give this value as $(\rho +p)/c^2$, equation (\ref{eq:rmassin}) provides  a different value $(\rho +3p)/c^2$.
Perhaps the origin of the problem is not in the gravitational energy but in $T^{ik}$ itself.

\bigskip
\noindent
$\blacktriangleright$
Given this backdrop, it thus appears that the relativistic formulation of matter given by $T^{ik}$ suffers from some subtle inherent problems. The point to note is that there is no role of the notorious (pseudo) energy of the gravitational field in these problems.
It would not be correct to conclude that the above-analysis advocates denial of fluid pressure in GR (as the problems are evaded in the absence of pressure).
Rather it insinuates that the 4-dimensional description of matter in terms of $T^{ik}$ is not compatible with the geometric description of gravitation.
It is perhaps not correct to patchwork a 4-dimensional tensor from two basically distinct kinds of 3-dimensional quantities  $-$ (i) the energy density, a non-directional quantity and (ii) the momenta and stresses, directional quantities. The tensor however treats them at equal footing by recognizing a component $T^{ik}$ as a scalar (irrespective of the values of $i$ and $k$) linked with the surface specified by $i$ and $k$ in the {\it hypothetical} 4-dimensional fluid, in the same way as the component $G^{ik}$ is linked with the curvature of the same surface. This leads to a sound mathematics and we do not notice any inconsistency until we relate the tensor $T^{ik}$ with the {\it real}  fluid which is 3-dimensional and {\it not 4-dimensional}.

Does it then mean that Einstein's `wood' is not only low grade compared to the standards
of his `marble' but it is also infested? It should be noted that the relativistic
formulation of the matter, in terms of the tensor $T^{ik}$,
has never been tested in any direct experiment.
It may be recalled that the crucial tests of GR, which have substantiated the theory beyond doubt, are based on the solutions of equation (\ref{eq:RicciEq}) only, viz. the Schwarzschild and Kerr solutions.  

It thus becomes increasingly clear that the development of GR was led astray by formulating matter in terms of $T^{ik}$. This is corroborated by the fact that whenever the theory takes recourse  to $T^{ik}$ in equation  (\ref{eq:EinsteinEq}), the trouble shows up in the form of either the dark energy or the inviabilities of Godel's solution and Schwarzschild's interior solution, etc.
In view of the new finding, this assertion acquires a new meaning - we have been searching for the matter in the wrong place. The correct place to search for it is the geometry. We have seen in innumerable examples that matter is already present in the geometry of equation (\ref{eq:RicciEq}) without taking recourse to $T^{ik}$. That is, the `wood' is already included into the `marble', dramatically fulfilling Einstein's obsession!

\section{Successes of the Novel Gravity Formulation}

\subsection{Observational Support to the New Paradigm}

The last words on a putative theory has to be spoken by observations and experiments.
The consistency of the field equation (\ref{eq:RicciEq}) with the local observations in the solar system and binary pulsars, has already been established in the standard tests of GR - the only satisfactory testimonial of the theory among the conventional tests, which do not require any epicycle of the dark sectors.

Interestingly, as has been shown recently \cite{vishwa_milne}, all the cosmological observations can also be explained successfully in terms of a   homogeneous, isotropic solution of equation (\ref{eq:RicciEq}). This solution can be obtained by solving  (\ref{eq:RicciEq}) for the Robertson-Walker metric, yielding
\be
ds^2=c^2 dt^2-c^2t^2\left(\frac{dr^2}{1+r^2}+r^2d\theta^2+r^2\sin^2\theta ~d\phi^2\right),\label{eq:milne}
\ee
which represents the homogeneous, isotropic Universe in the new paradigm. It may be mentioned that solution (\ref{eq:milne}) (which is generally recognized as the Milne model) wherein the Universe appears dynamic in terms of the comoving coordinates and the cosmic time $t$, can be reduced to the Minkowskian form by using the locally defined measures of space and time \cite{vishwa_milne}. 

The observational tests considered in \cite{vishwa_milne} include the observations of the high-redshift supernovae (SNe) Ia, the observations of high-redshift radio sources, observations of starburst galaxies,  the CMB observations and
compatibility of the age of the Universe with the oldest objects in it (for instance, the globular clusters) for the currently measured values of the Hubble parameter.
It may also be mentioned that by preforming a rigorous statistical test on a much bigger sample of SNe Ia
(by taking account of the empirical procedure by which corrections are made to their absolute magnitudes), a recent study  has found only a marginal evidence for an accelerated expansion and the 
data are quite consistent with the Milne model \cite{Sarkar}.

One may wonder how the new model, which does not possess dark energy (and hence not an accelerated expansion), manages to reconcile with the observations. The mystery lies in the special expansion dynamics of the model at a constant rate throughout the evolution, as is clear from  
equation (\ref{eq:milne}), wherein the Robertson-Walker scale factor $S=ct$. We note that, unlike the standard cosmology, solution (\ref{eq:milne}) efficiently provides different measures of distances without requiring any input from the matter fields. For instance, the luminosity distance $d_{\rm L}$ of a source of redshift $z$, in the present case, is given by
\be
d_{\rm L}=cH_0^{-1}(1+z)\sinh[\ln(1+z)],\label{eq:d_L}
\ee
where $H_0$ represents  the present value of the Hubble parameter $H=\dot{S}/S$. As has been shown in Figure 1, the luminosity distance of an object of redshift $z$ in the new cosmology is almost the same as that in the standard cosmology for $z\simlt 1.3$. This explains why both models are equally consistent with the SNe Ia data wherein the majority of the SNe belong to this range of redshift. However, for $z> 1.3$, the new model departs significantly from the standard cosmology, as is clear from the figure. Hence observations of more SNe Ia at higher redshifts will be decisive for both paradigms.

\begin{figure}
\begin{center}
\resizebox{9.5cm}{!}{\includegraphics{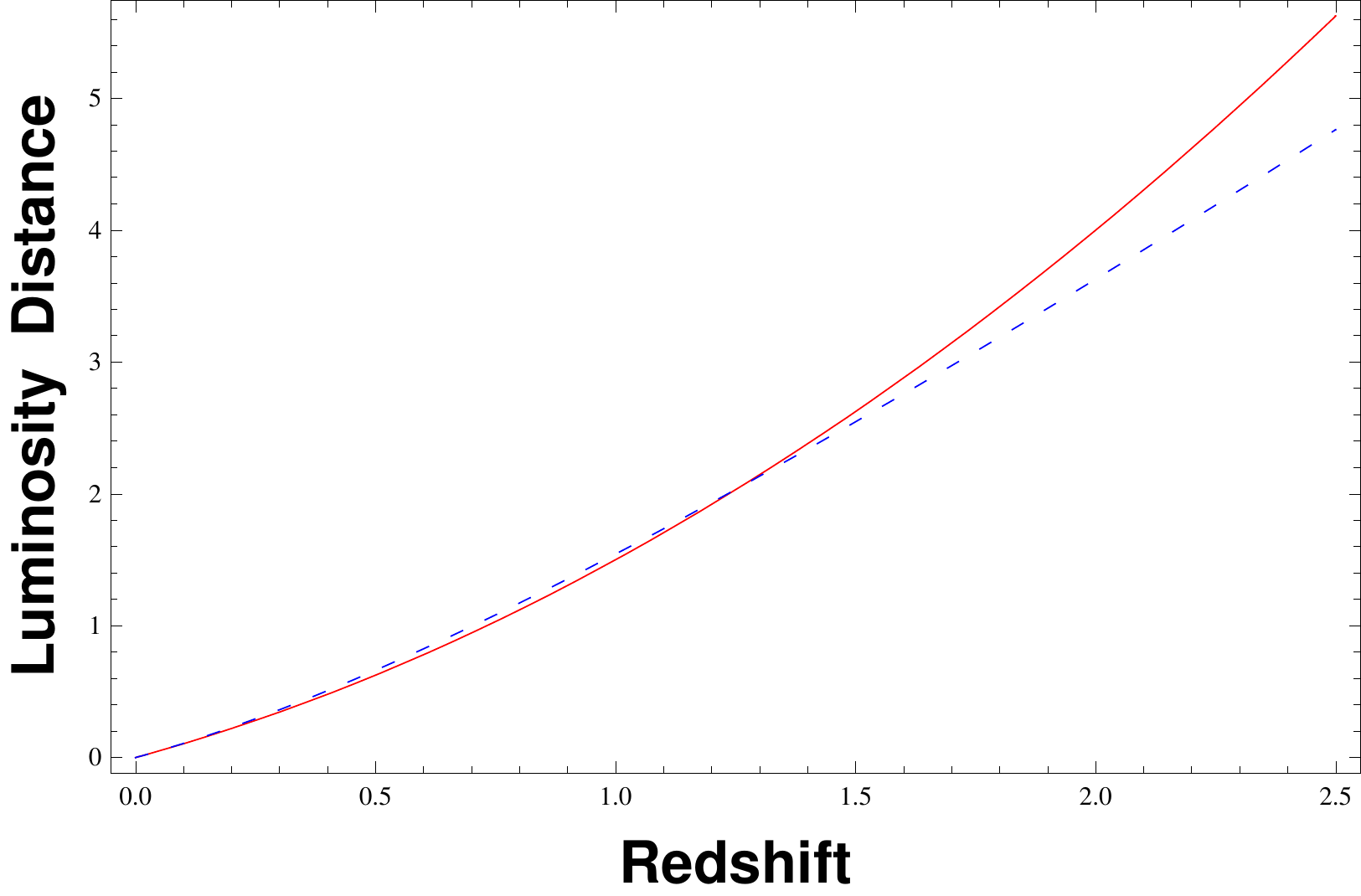}}
\end{center}
{\caption{\small Luminosity distance in the new model (continuous curve) is compared with that in the $\Lambda$CDM concordance model $\Omega_{\rm m}=1-\Omega_\Lambda=0.3$ (broken curve). Distances shown on the vertical axis are measured in units of $cH_0^{-1}$. The two models significantly depart for $z\simgt 1.3$.
}}
\end{figure}

\subsection{Different Pieces Fit Together}

\medskip
\noindent
$\blacktriangleright$
As the dark energy can be assimilated in the energy-stress tensor, and since the latter is absent from the dynamical equations in the new paradigm (wherein the fields appear through the geometry), the dark energy and its associated problems, for instance the  cosmological constant problem (which appears due to a conflict between the energy-stress
tensor $T^{ik}$ in equation (\ref{eq:EinsteinEq}) and the energy density of vacuum in the quantum field theory) and the coincidence problem, are evaded in the new paradigm.

For the same reason is circumvented the flatness problem due to the absence of $T^{ik}$ in the new paradigm.

\medskip 
\noindent
$\blacktriangleright$
As has been mentioned earlier, the observed isotropy of CMB cannot be explained in the standard paradigm in terms of some homogenization process taken place in the baryon-photon plasma operating under the principle of causality, since a finite value for the particle horizon    
$
d_{\rm PH}(t)=c S(t) \int^{t}_0 dt'/S(t')
$
(the largest distance from which light could have reached the present observer) exists in the theory. As $d_{\rm PH}=\infty$ always for  $S=ct$, no horizon exists  in the new paradigm and the whole Universe is always causally connected, which explains the observed overall uniformity of CMB without invoking inflation \cite{FrontPhys}.

\medskip
\noindent
$\blacktriangleright$
As the Big Bang singularity is a breakdown of the laws of physics and the geometrical structure of spacetime,  there have been attempts to discover singularity-free cosmological solutions of Einstein equations, which is usually achieved by violating the energy conditions.

Although solution (\ref{eq:milne}), which represents the cosmological model in the new paradigm, has well-behaved metric potentials at $t=0$,  the volume of the spatial slices vanishes there resulting in a blowup in the accompanied matter density. However, this is just an illusory coordinate effect which can be removed in the Minkowskian form of solution (\ref{eq:milne}) by considering the locally defined  coordinates of space and time. 

Moreover, as the locally defined time scale $\tau$ is related with the cosmic time $t$ through the transformation $\tau=t_0 \ln(t/t_0)$ \cite{vishwa_milne},  the epoch corresponding to the big bang, is pushed back to the infinite past giving an infinite age to the Universe which can accommodate  even older objects than the standard cosmology can.
Interestingly, even in terms of the cosmic time $t$, wherein the Universe appears dynamic, the age of the Universe appears higher than that in the standard paradigm \cite{vishwa_milne}.

\medskip
\noindent
$\blacktriangleright$
As has been mentioned earlier, the conventional `source term' $T^{ik}$ in equation (\ref{eq:EinsteinEq}) fails to include the energy, momentum or angular momentum of the gravitational field. Remarkably, these quantities, akin to the matter fields, are inherently present in the geometry of equation (\ref{eq:RicciEq}), substantiating the new strategy of the new paradigm to represent the source through geometry.
For instance, the term $K/r=-2Gm/(c^2r)$ in the Schwarzschild solution (\ref{eq:Sch}) contains the gravitational energy at the point $r$. It perfectly agrees with the Newtonian estimate of the gravitational energy given by $-Gm/r$ indicating that the term $-2Gm/(c^2r)$ is just its relativistic analogue.
Assigning the gravitational energy to $K/r$, is also supported by the locality of GR, which becomes an intrinsic characteristic of the theory as soon as the Newtonian concept of gravitation as a force (action-at-a-distance) is superseded by the curvature. Being a local theory, GR then assigns the curvature present at a particular point, to the source present at that very point. Thus, the agent responsible for the curvature  in (\ref{eq:Sch}) must be the gravitational energy, since matter exists only at $r=0$ whereas (\ref{eq:Sch}) is curved at all finite values of $r$. 
Hence the presence of curvature in the Schwarzschild solution implies that the gravitational energy does gravitate just as does every other form of energy, and the gravitational field is obviously present in the geometry of equation (\ref{eq:RicciEq}).

Similarly the angular momentum of the gravitational field, arising from the rotation of the mass $m$, is revealed through the geometry of the Kerr solution, and its momentum in the Taub-NUT solution.
Thus the long-sought-after gravitational field energy-momentum-angular momentum of GR is already present in the geometry.

\medskip
\noindent
$\blacktriangleright$
It may be interesting to note that new interior solutions, based on the solutions of  equation (\ref{eq:RicciEq}), have been formulated in the new paradigm which form the Schwarzschild interior and the Kerr interior \cite{IJGMMP}. The new interiors are conceptually satisfying and free from the earlier mentioned problems.

\medskip
\noindent
$\blacktriangleright$
As the Newtonian theory of gravitation provides excellent approximations under a wide range of astrophysical cases, the first crucial test of any theory of gravitation is that it reduces to the Newtonian gravitation  in the limit of a weak gravitational field. In this context, it has been recently shown \cite{IJGMMP} that the new paradigm consistently admits the Poisson equation in the case of a slowly varying weak gravitational field when the concerned velocities are considered much less than c, provided we take into account the inertial as well as the gravitational properties of matter, as should correctly be expected in a true Machian theory\footnote{ 
The standard paradigm on the other hand fails to fulfill this requirement as equation (\ref{eq:EinsteinEq}), in the limit of the weak field, does not reduce to the Poisson equation in the presence of a non-zero $\Lambda$ (or
any other candidate of dark energy), which becomes unavoidable in the standard paradigm. Also, it would not be correct to argue that a  $\Lambda$ as small as $\approx 10^{-56}$ cm$^{-2}$ (as inferred from the cosmological observations) cannot contribute to the physics appreciably in the local problems. It has been shown recently that even this value of $\Lambda$ does indeed contribute to the bending of light and to the advance of the perihelion of planets \cite{Ishak}.}.

\medskip
\noindent
$\blacktriangleright$
Interestingly, the new paradigm becomes scale invariant, since the new field equation (\ref{eq:RicciEq}) is manifestly scale invariant. This becomes a remarkable achievement in the sense that one of the most common ways for a theory, with continuous field, to be renormalizable is for it to be scale invariant.

\medskip
\noindent
$\blacktriangleright$
Since the Universe in the new paradigm is flat, the symmetries of its Minkowskian form make it possible to validate the conservation of  energy, solving the long-standing problems associated with the conservation of energy. As has been shown by Noether, it is the symmetry of the Minkowskian space which is the cause of the conservation of the energy momentum of a physical field \cite{Noether, Baryshev}.

\subsection{Geometrization of Electromagnetism in the New Paradigm}

How can the electromagnetic field be added to the new paradigm?
While the equivalence principle renders the gravitational and inertial fields essentially geometrical (owing to the fact that the ratio of the gravitational and inertial mass is strictly unity for all matter), this is not so in the case of the electromagnetic field (since the ratio of electric charge to mass varies from particle to particle). Hence, the addition of the electromagnetic energy tensor $E^{ik}$  to equation (\ref{eq:EinsteinEq}), results in
\be
R^{ik}=-\frac{8\pi G}{c^4}E^{ik},\label{eq:e-m-corr}
\ee
since  $T^{ik}$ is absorbed in the geometry (as we have noted earlier), and  $g_{ik} E^{ik}=0$ reduces  $R=0$ identically.
The tensor $E^{ik}$  is given, in terms of the skew-symmetric electromagnetic field tensor $F_{ik}$, as usual:
\be
E^{ik}=\nu\big[-g^{k\ell} F^{ij}F_{\ell j}+\frac{1}{4}g^{ik} F_{\ell j}F^{\ell j}\big],\label{eq:e-m}
\ee
where $\nu$ is a constant. It has already been shown that  equation (\ref{eq:e-m-corr}), taken together with the `source-free' Maxwell's equations
\be\label{eq:Max-corr}
\left.\begin{aligned}
\frac{\partial F_{ik}}{\partial x^\ell}+\frac{\partial F_{k \ell}}{\partial x^i}+\frac{\partial F_{\ell i}}{\partial x^k}=0\\
\frac{\partial}{\partial x^k}(\sqrt{-g}F^{ik})=0
 \end{aligned}
 \right\},~~ g={\rm det}((g_{ik})),
\ee
consistently represents electromagnetic field in the presence of gravitation \cite{IJGMMP}. As the existence of charge is intimately related with the existence of the charge-carrier matter and since the new paradigm claims the inherent presence of matter in the geometry, it is reasonable to expect the charge also to appear through the geometry. This view is indeed supported not only by the Reissner-Nordstrom and Kerr-Newman solutions, but also by the cosmological solutions - the so-called `electrovac universes'\footnote{Let us note that unlike the Reissner-Nordstrom and Kerr-Newman solutions (which represent the field outside the charged matter), the electrovac solutions are not expected to contain any `outside' where the charge-carrier matter can exist.}  \cite{IJGMMP}, wherein the charge does appear through the geometry.

Thus, equations (\ref{eq:e-m-corr}) - (\ref{eq:Max-corr}) of restricted validity in the standard paradigm [wherein they are believed to represent the electromagnetic field in vacuum, very much in the same vein as equation (\ref{eq:RicciEq}) is believed to represent the gravitational field in vacuum] get full validity and represent an unified theory of gravitation, inertia and electromagnetism. 

Interestingly, Misner and Wheeler also expressed similar views long ago and advocated to represent {\it ``gravitation, electromagnetism, unquantized charge and unquantized mass as properties of curved empty space''} \cite{Misner-Wheeler}. Although they failed to realize the presence of fields in the flat spacetime\footnote{The removal of charge (by switching off $E^{ik}$ from (\ref{eq:e-m-corr}), in which case the`electrovac universes' become flat) does not  mean that mass (which was carrying charge) must necessarily disappear from these solutions.}, nonetheless they too realized that equations (\ref{eq:e-m-corr}) - (\ref{eq:Max-corr}) provide  a unified theory of electromagnetism and gravitation.

\section{Summary and Conclusions: What Next?}

GR is undoubtedly a theory of unrivaled elegance. The theory indoctrinates that gravitation is a manifestation of the spacetime geometry - one of the most precious insights in the history of science.
It has emerged as a highly successful theory of gravitation and cosmology, predicting several new phenomena, most of them have already been confirmed by observations. The theory has passed every observational test ranging from the solar system to the largest scale, the Universe itself.

Nevertheless, GR  ceases to be the ultimate description of gravitation, an epitome of a perfect theory, despite all these feathers in its cap. Besides its much-talked-about incompatibility with quantum mechanics, the theory suffers from many other conceptual problems (discussed in the preceding sections), most of which are generally ignored.  If in a Universe where, according to the standard paradigm, some 95\% of the total content is still missing, it is an alarming signal for us to turn back to the very foundations of the theory.
In view of these problems, we are led to believe that the historical development of GR was indeed on a wrong track and the theory requires modification or at least reformulation.

By a critical analysis of Mach's principle and the equivalence principle, a new insight with deeper vision of  a geometric theory of gravitation
emerges:  matter, in its entirety of gravitational, inertial and electromagnetic properties, can be fashioned out of spacetime itself.
This  revolutionizes our views on the representation of the source of curvature/gravitation by dismissing the conventional source representation through the energy-stress tensor of matter $T^{ik}$ and establishing the spacetime itself as the source.

This appears as the missing link of the theory and posits that spacetime does not exist without matter, the former is just an offshoot of the latter. The conventional assumption that matter only fills the already existing spacetime, does not seem correct. 
This establishes the canonical equation $R^{ik}=0$ as the field equation of gravitation plus inertia in the very presence of matter, giving rise to a new paradigm in the framework of GR.
Though there seems to exist some emotional resistance in the community to tinkering with the elegance of GR, however the new paradigm dramatically enhances the beauty of the theory in terms of the deceptively simple new field equation $R^{ik}=0$.
Remarkably the new paradigm explains the observations at all scales without requiring the epicycle of dark energy.

This review provides an increasingly clear picture that the new paradigm is a viable possibility in the framework of GR, which is valid at all scales, avoids the fallacies, dilemmas and paradoxes, and answers the questions that the old framework could not address.

Though we have witnessed numerous evidences of the presence of fields in the solutions of the field equation (\ref{eq:RicciEq}), however,
the challenge to discover, from more fundamental considerations, a concrete mathematical formulation of the fields in purely geometric terms is still to be met.
 This formulation is expected to use the gravito-electromagnetic features of GR in the new paradigm and is expected to achieve the following:

\begin{enumerate}

\item
It should explain the observed flat rotation curves of galaxies without requiring the ad-hoc dark matter.

\item
The net field in a homogeneous and isotropic background must be vanishing.

\end{enumerate}

\acknowledgments{Acknowledgment} The author thanks IUCAA for hospitality where part of this work was done during a visit. Thanks are also due to two anonymous referees, to one for making critical, constructive comments which helped improving the manuscript; and to other for pointing out an old work of Misner and Wheeler \cite{Misner-Wheeler}, which is deeply connected with and strongly supporting the present work.

\bibliographystyle{mdpi}
\makeatletter
\renewcommand\@biblabel[1]{#1. }
\makeatother

\end{document}